\documentclass[sigconf]{acmart}
\usepackage[whole]{bxcjkjatype}
\usepackage{bm}

\AtBeginDocument{%
  \providecommand\BibTeX{{%
    \normalfont B\kern-0.5em{\scshape i\kern-0.25em b}\kern-0.8em\TeX}}}

\setcopyright{rightsretained} 
\copyrightyear{2022} 
\acmYear{2022} 
\acmConference{SA '22 Posters}{December 06-09, 2022}{Daegu, Republic of Korea}\acmBooktitle{SIGGRAPH Asia 2022 Posters (SA '22 Posters), December 06-09, 2022}\acmDOI{10.1145/3550082.3564190}
\acmISBN{978-1-4503-9462-8/22/12}

\begin{document}

\title{Accelerated and Optimized Search of Imperceptible Color Vibration for Embedding Information into LCD images}

\author{Shingo Hattori}
\affiliation{%
  \institution{University of Tsukuba}
  \streetaddress{}
   \city{}
   \state{}
   \country{}
   \postcode{}
}
\email{shatto@pml.slis.tsukuba.ac.jp}

\author{Takefumi Hiraki}
\affiliation{%
  \institution{University of Tsukuba}
   \streetaddress{}
   \city{}
   \state{}
   \country{}
   \postcode{}
}
\email{hiraki@slis.tsukuba.ac.jp}

\renewcommand{\shortauthors}{S. Hattori and T. Hiraki.}





\begin{teaserfigure}
  \centering
  \includegraphics[width=0.92\textwidth]{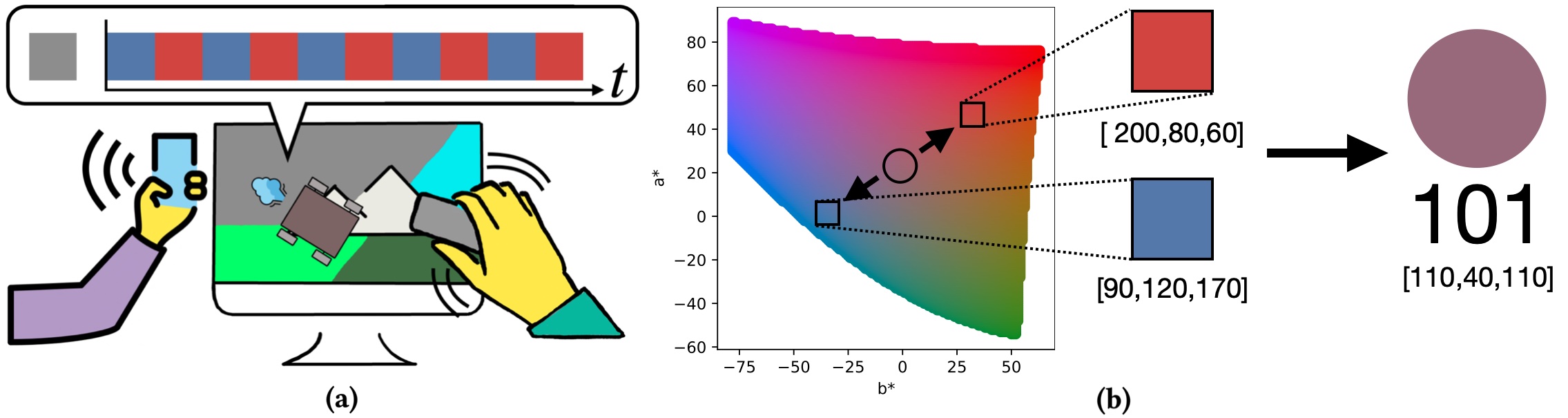}
  \caption{(a): Concept of information embedding on images using imperceptible color vibration. Devices can detect areas that appear gray to humans as switching between red and blue and acquire information on the presence of this switching to perform various actions.
  (b): Overview of a search method of imperceptible color vibration. Color pairs are chosen in a L*a*b* color space with fixed luminance (L*), and the difference in RGB values between color pairs determines the presence of color vibration.}
  \label{fig:teaser}
\end{teaserfigure}
\maketitle

\section{Introduction}
Large, high-resolution displays are installed throughout the city as public displays.
By superimposing invisible information on the images of these displays, large numbers of devices with cameras and sensors can communicate with the displays without prior pairing.
Several applications have been proposed, such as operating robots or communicating information to users by displaying 2D codes on images~\cite{aColiseum}.
However, the display of 2D codes has the problem of compromising the appearance of displayed content.


Abe~\textit{et al.}~\cite{abesa} proposed a method of communicating with devices by superimposing invisible information using color vibration on images displayed on off-the-shelf liquid-crystal displays (LCD).
Using this method, we can embed the information for devices in images without interfering with the displayed content (Fig.~\ref{fig:teaser}a).

Abe~\textit{et al.}~\cite{abesa} uses a simple serial loop operation to search for color pairs comprising a color vibration, which requires a very long processing time due to the huge search space.

\begin{figure*}[t]
  \centering
  \includegraphics[width=0.85\linewidth]{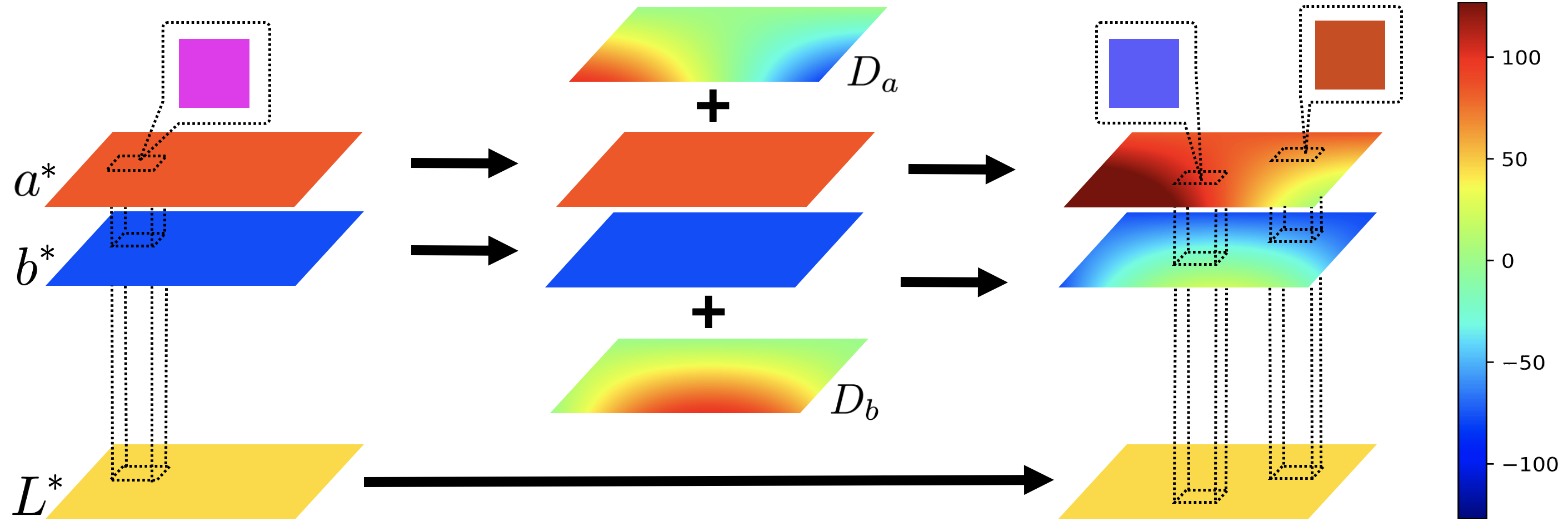}
  \caption{Overview of the proposed search method of imperceptible color vibration. We decompose the target color (purple) into a three-channel array of L*, a*, and b* channels. We add the parameter arrays $\bm D_a$ and $\bm D_b$, which represent movement in color space, to the a* and b* channel arrays, respectively. We can obtain multiple candidate combinations of color pairs by recombining the L* channel array with the a* and b* channel arrays.}
  \label{fig:search}
\end{figure*}

In this paper, we propose an accelerated and optimized search method for color pairs that constitute the imperceptible color vibration for embedding information on LCD images.
To achieve fast color pair search, we parallelized the search process, which is previously done individually, by using arrays representing the amount of movement and an operation to extract elements from the array that satisfy the conditions.
In addition, we investigate the amount of information that can be superimposed on nine color images using the imperceptible color vibration and clarify the applicability of embedding information into images using the color vibration.

\section{Methods}
Information embedding by the imperceptible color vibration utilizes the characteristic that when the vibration frequency of two colors with fixed luminance exceeds the critical color fusion frequency (CCFF) of approximately 25 Hz, humans perceive the fused color~\cite{jiang2007human}.
The color pairs used in the imperceptible color vibration are chosen to be two colors that are point-symmetric with respect to the target color in a L*a*b* color space with fixed luminance (Fig.~\ref{fig:teaser}b).

In the proposed method, we generate an array composed of a set of color pairs and extract the color pairs that satisfy the color vibration condition (cannot see by humans) and the embedding signal condition (can detect the vibration by devices).

Figure~\ref{fig:search} shows the overview of the proposed search method.

We first decompose the target color into the L*, a*, and b* channel components and create arrays that store these values.



We add the arrays $D_a (D_{aij} := i \sin j)$ and $D_b (D_{bij} := i \cos j)$ to a* array and b* array.
$D_a$ and $D_b$  represent the amount of movement on a* axis and b* axis in a L*a*b* color space with fixed luminance 
shown as Fig.~\ref{fig:teaser}b.
Then, we recombine the added arrays with the L* channel array to create an array of colors that have been shifted in color space from the target color.
Similarly, we can create a point-symmetric color pair about the target color by subtracting $D_a$ and $D_b$ from a* array and b* array.
Thanks to these processes, the proposed method enables an efficient search for color pairs in parallel by using matrix calculations.

\section{Experiments}
\begin{table}[b]
\caption{Colors, embedding information, and threshold values used in the colour pair search.}
\scalebox{0.96}{
\begin{tabular}{|c|p{55mm}|}
\hline
Colors (RGB) & Black (100, 100, 100), Gray (170, 170, 170), White (240, 240, 240), Red (240, 100, 100), Green (100, 240, 100), Blue (100, 100, 240), Yellow (240, 240, 100), Cyan (100, 240, 240), Magenta (240, 100, 240) \\ \hline
Information (RGB) 
& 100, 010, 001, 110, 101, 011, 111 \\ \hline
$V_{th}$ & 50, 100, 150, 200 \\ \hline
$R_{novib}$ & 0.5, 0.25, 0.125 \\ \hline
\end{tabular}
}
\label{tab:search_params}
\end{table}

We implemented the search for imperceptible color vibration using the previous~\cite{abesa} and proposed methods.
We explored color vibrations using a PC (MacBook Pro 13-inch, 2019, CPU: Intel Core i5-8279U, Memory: 16 GB) for the parameter and color combinations listed in Table~\ref{tab:search_params} and compared the time taken to perform them.
We define a bit as high (1) if the change in the three channels (RGB) in which vibrations can be embedded in color vibration exceeds provisional perception threshold $V_{th}$, and low (0) if the change is below $V_{th} \times R_{novib}$.

The time required for the search was 3423.07 s for the previous method and 62.62 s for the proposed method, which means that the proposed method was about 54 times faster.
Table~\ref{tab:embed_pattern} shows the results of the search for combinations of signals that can be embedded in representative nine-color images.
For the six colors with small G values (black, gray, white, red, blue, and magenta), it was found that more than four types of information can be superimposed.

\begin{table}[t]
       \centering
       \caption{The presence of color pairs that satisfy the conditions for each color and embedded information.}
       \scalebox{0.69}{
        \begin{tabular}{|c|c|c|c|c|c|c|c|c|c|}
        \hline
        \begin{tabular}{c}
        Information\\ (RGB) 
        \end{tabular}
        & Black & Gray & White & Red & Green & Blue & Yellow & Cyan & Magenta  \\ \hline
        100   & \checkmark  & \checkmark  & \checkmark  &    & \checkmark  & \checkmark  & \checkmark  & \checkmark  & \checkmark  \\ \hline
        010   &  &   &    &   &    &    &    &    &   \\ \hline
        001   & \checkmark  & \checkmark  & \checkmark  & \checkmark  & \checkmark  & \checkmark  & \checkmark  & \checkmark  & \checkmark  \\ \hline
        110   &   & \checkmark  & \checkmark  & \checkmark  &    &   &    &    & \checkmark  \\ \hline
        101   & \checkmark  & \checkmark  & \checkmark  & \checkmark  & \checkmark  & \checkmark  & \checkmark  & \checkmark  & \checkmark  \\ \hline
        011   &  &  &    &  &    &    &    &    &    \\ \hline
        111   & \checkmark  & \checkmark  & \checkmark  & \checkmark  &    & \checkmark  & \checkmark  &    & \checkmark  \\ \hline
        \end{tabular}
        }
       \label{tab:embed_pattern}
\end{table}

\section{Conclusion}
We proposed a high-speed and optimized search method for imperceptible color vibration for embedding information in LCD images.
In future work, we will evaluate whether the color vibration is perceived as the same as the original color by user experiments.

\section*{Acknowledgements}
This work was supported by JST ACT-X Grant Number JPMJAX190O.

\bibliographystyle{ACM-Reference-Format}
\bibliography{ref} 


\begin{thebibliography}{3}


\ifx \showCODEN    \undefined \def \showCODEN     #1{\unskip}     \fi
\ifx \showDOI      \undefined \def \showDOI       #1{#1}\fi
\ifx \showISBNx    \undefined \def \showISBNx     #1{\unskip}     \fi
\ifx \showISBNxiii \undefined \def \showISBNxiii  #1{\unskip}     \fi
\ifx \showISSN     \undefined \def \showISSN      #1{\unskip}     \fi
\ifx \showLCCN     \undefined \def \showLCCN      #1{\unskip}     \fi
\ifx \shownote     \undefined \def \shownote      #1{#1}          \fi
\ifx \showarticletitle \undefined \def \showarticletitle #1{#1}   \fi
\ifx \showURL      \undefined \def \showURL       {\relax}        \fi
\providecommand\bibfield[2]{#2}
\providecommand\bibinfo[2]{#2}
\providecommand\natexlab[1]{#1}
\providecommand\showeprint[2][]{arXiv:#2}

\bibitem[Abe et~al\mbox{.}(2017)]%
        {abesa}
\bibfield{author}{\bibinfo{person}{S. Abe}, \bibinfo{person}{A. Arami},
  \bibinfo{person}{T. Hiraki}, \bibinfo{person}{S. Fukushima}, {and}
  \bibinfo{person}{T. Naemura}.} \bibinfo{year}{2017}\natexlab{}.
\newblock \showarticletitle{Imperceptible Color Vibration for Embedding
  Pixel-by-Pixel Data into LCD Images}. In \bibinfo{booktitle}{\emph{Proc. of
  CHI 2017 EA}}. \bibinfo{pages}{1464--1470}.
\newblock


\bibitem[Jiang et~al\mbox{.}(2007)]%
        {jiang2007human}
\bibfield{author}{\bibinfo{person}{Y. Jiang}, \bibinfo{person}{K. Zhou}, {and}
  \bibinfo{person}{S. He}.} \bibinfo{year}{2007}\natexlab{}.
\newblock \showarticletitle{Human visual cortex responds to invisible chromatic
  flicker}.
\newblock \bibinfo{journal}{\emph{Nature neuroscience}} \bibinfo{volume}{10},
  \bibinfo{number}{5} (\bibinfo{year}{2007}), \bibinfo{pages}{657--662}.
\newblock


\bibitem[Sugimoto et~al\mbox{.}(2005)]%
        {aColiseum}
\bibfield{author}{\bibinfo{person}{M Sugimoto}, \bibinfo{person}{G Kagotani},
  \bibinfo{person}{M Kojima}, \bibinfo{person}{H Nii}, \bibinfo{person}{A
  Nakamura}, {and} \bibinfo{person}{M Inami}.} \bibinfo{year}{2005}\natexlab{}.
\newblock \showarticletitle{Augmented coliseum: Display-based computing for
  augmented reality inspiration computing robot}. In
  \bibinfo{booktitle}{\emph{Proc. of SIGGRAPH 2005 Emerging Technologies}}.
  \bibinfo{pages}{1}.
\newblock


\end{thebibliography}

\end{document}